\documentclass[nofootinbib, a4paper, showkeywords, showpacs,superscriptaddress,groupedaddress,preprint]{revtex4}

\usepackage{amsmath,mathrsfs,amscd,graphicx,cancel}
\usepackage{multirow}
\usepackage{color}
\usepackage{rotating}
\usepackage{cancel}
\usepackage{subfig}
\usepackage{textcomp}
\usepackage[final]{pdfpages}
\usepackage{array}
\usepackage{float}
\usepackage{url}
\usepackage{textcomp}
\usepackage{amsmath}
\usepackage{amsfonts, latexsym, epsfig}
\usepackage{bm}
\usepackage{times}
\usepackage{epsfig}
\usepackage{amssymb}
\usepackage{hyperref}

\def\beq{\begin{equation}}
\def\eeq{\end{equation}}
\def\bea{\begin{eqnarray}}
\def\eea{\end{eqnarray}}

\begin{document}
\title{Quark jet versus gluon jet: fully-connected neural networks with high-level features}
\author{Hui Luo}
\email{hluo@gscaep.ac.cn}
\affiliation{Graduate School of China Academy of Engineering Physics, Beijing, 100193, China}
\affiliation{Zhejiang Institute of Modern Physics and Department of Physics, Zhejiang University, Hangzhou, Zhejiang 310027, China}
\affiliation{II. Institut f\"ur Theoretische Physik, Universit\"at Hamburg\\ Luruper Chaussee 149, D- 22761 Hamburg, Germany}
\author{Ming-xing Luo}
\affiliation{Zhejiang Institute of Modern Physics and Department of Physics, Zhejiang University, Hangzhou, Zhejiang 310027, China}
\author{Kai Wang}
\affiliation{Zhejiang Institute of Modern Physics and Department of Physics, Zhejiang University, Hangzhou, Zhejiang 310027, China}
\author{Tao Xu}
\affiliation{Zhejiang Institute of Modern Physics and Department of Physics, Zhejiang University, Hangzhou, Zhejiang 310027, China}
\author{Guohuai Zhu}
\affiliation{Zhejiang Institute of Modern Physics and Department of Physics, Zhejiang University, Hangzhou, Zhejiang 310027, China}

\begin{abstract}
Jet identification is one of the fields in high energy physics that machine learning has begun to make an impact. More often than not, convolutional neural networks are used to classify jet images with the benefit that essentially no physics input is required. Inspired by a recent work by Datta and Larkoski, we study the classification of quark/gluon-initiated jets based on fully-connected neural networks (FNNs), where expert-designed physical variables are taken as input. FNNs are applied in two ways: trained separately on various narrow jet transverse momentum $p_{TJ}$ bins; trained on a wide region of $p_{TJ} \in [200,~1000]$ GeV. We find their performances are almost the same. The performance is better when the $p_{TJ}$ is larger. Jet discrimination with FNN is studied on both particle and detector level data. The results based on particle level data are comparable with those from deep convolutional neural networks, while the significance improvement characteristic (SIC) from detector level data would at most decrease by $15\%$.  We also test the performance of FNNs with full set or subsets of jet observables as input features. The FNN with one subset consisting of fourteen observables shows nearly no degradation of performance. This indicates that these fourteen expert-designed observables could have captured the most necessary information for separating quark and gluon jets.
\end{abstract}


\maketitle

\section{Introduction}
At the Large Hadron Collider (LHC), hadronic decay final states in processes including $W/Z$ bosons or squarks in supersymmetric theories are dominated by light-quark-initiated jets, while the corresponding Standard Model~(SM) background often consists of gluon-initiated jets.
This indicates the importance of discrimination between quark jets and gluon jets to optimize background analysis for new physics searches.
It has been known that the two kinds of jets are qualitatively different since early measurements at PETRA and LEP colliders.
For instance, radiation from color octet gluon will result in a jet of larger width compared to the one from quark radiation.
However, the qualitative features distinguishing quark jets from gluon jets were never as robust as the well-known b-jet tagging until the practical jet tagging via charged particle multiplicity and jet width proposed in \cite{Gallicchio:2011xq,Gallicchio:2012ez} was finally employed by the ATLAS collaboration.
Besides these global features, local information observables such as N-subjettiness~\cite{Stewart:2010tn,Thaler:2010tr,Thaler:2011gf}, energy correlation functions~\cite{Larkoski:2013eya,Moult:2016cvt}, etc., are also used to distinguish different jet species. The idea of jet substructure is recently reviewed in \cite{Larkoski:2017jix, Gras:2017jty}.

At this stage, one challenging task is to overcome the difficulties in performing analysis on the large and high-dimensional jet datasets. For this purpose, there's a growing interest in exploring the potential of machine learning in high energy physics studies. Shallow neural networks, as an example, have been used for various purposes for a long time. The deep neural network technology also made impressive breakthrough in pattern recognition.
Recently, these deep neural networks have been applied in searches for new physics~\cite{Baldi:2014kfa, Baldi:2016fzo}, the identification of boosted jets~\cite{Baldi:2014pta,Almeida:2015jua,Searcy:2015apa,Behr:2015oqq,Conway:2016caq,Santos:2016kno,Alves:2016htj,Kasieczka:2017nvn, Pearkes:2017hku,Erdmann:2017hra,Butter:2017cot}, general jet classifications~\cite{Baldi:2016fql,Guest:2016iqz,deOliveira:2015xxd,Komiske:2016rsd,Louppe:2017ipp,Shimmin:2017mfk,Datta:2017rhs,Datta:2017lxt,Cheng:2017rdo}, neutrino physics studies~\cite{Racah:2016gnm,Aurisano:2016jvx,Renner:2016trj} and other related topics~\cite{Barnard:2016qma,Pang:2016vdc,deOliveira:2017pjk,Sadowski:2017ilo,Cohen:2017exh,Komiske:2017ubm,Erdmann:2017str,Metodiev:2017vrx,
Ren:2017ymm,Chang:2017kvc}.

Quark and gluon jet discrimination, being a typical classification problem, might be solved with these tools as well. Making use of the energy deposition in calorimeters, jet information is represented by grayscale image data and underlying features could be extracted by deep convolutional neural networks (DCNNs), which is proved to be very powerful in computer vision~\cite{deOliveira:2015xxd}. As an attempt to include more well-studied tagging observables such as charged particle multiplicity, the authors of \cite{Komiske:2016rsd} used pixel-level charged particle counts, transverse momentum $p_T$ of charged and neutral particles as three ``colors" of jet images. They found the DCNN already outperformed traditional methods. This result is very impressive, especially considering almost none of the expert-designed jet observables are used. Nevertheless, there's still discussion on whether DCNN is the optimal choice for jet classification~\cite{Pearkes:2017hku}. Unlike daily life pictures, jet images are often sparse, {\it i.e.} only few pixels are activated.
In the meantime, it's still difficult to see from the behaviors of DCNNs what physics is learned.

FNNs are intrinsically very effective in analyzing multidimensional problems. The input data for FNNs are jet observables rather than jet images. If FNNs could capture substructure features with only a finite number of input observables, the volume of dataset would be better under control. As a concrete example, it was demonstrated in \cite{Datta:2017rhs} that jet mass plus only eight N-subjettiness observables are enough to span the phase space of final states in the hadronic decay of boosted Z boson, in order to tell its differences from QCD jets. In this paper, we study the quark/gluon tagging using FNNs with different combinations of high-level observables. We find that jet mass plus N-subjettiness observables are not enough in the case of quark/gluon jet discrimination. Instead, our input features contain at most $36$ jet observables and the corresponding performance of the FNN is comparable to, or even slightly better in some regions than that of DCNNs. 
Another advantage of our method is that only one FNN is enough to separate quark/gluon jets with very different transverse momenta. This is not easy for DCNNs because jet images look different for various momenta.

In Sec.~\ref{sec:observables-and-data-generation}, we discuss jet observables used as the input of deep neural networks and the method to generate jet samples at parton level and detector level. In Sec.~\ref{sec:ArchitectureFNN} and Sec.~\ref{sec:results}, we discuss the architecture and the performance of our neural networks. In Sec.~\ref{sec:discussion},  we make a brief summary and discussion on the results.

\section{Observables and event generations}\label{sec:observables-and-data-generation}
In the first part of this section, we enumerate input observables to the neural networks, which elaborate both global and local properties of a jet.
As being pointed out in~\cite{Datta:2017rhs}, the choice of observable basis in the final state phase space is not unique.
We test different combinations of these features as input to weigh their importance.

The data generation process is discussed in detail in the second part of this section.
To generate a particle level dataset for training and testing neural networks, we simulate parton level events with MadGraph \cite{Alwall:2014hca} and parton shower process with PYTHIA \cite{Sjostrand:2014zea}. After that, final state particles are clustered with FastJet \cite{Cacciari:2011ma}, while observables are extracted using FastJet Contrib as well as our private codes.
We also generate a detector level dataset by adding an additional detector simulation step with Delphes \cite{deFavereau:2013fsa} before FastJet.

\subsection{Observables as inputs of neural networks}
Global jet observables are calculated with all jet constituents. We include full jet and charged particle multiplicities; jet mass, jet transverse momentum and their ratio; generalized angularities such as $p_T^D$, Les Houches Angularity, jet girth and thrust. On the other hand, local observables reconstructed with only part of the jet constituents also work because they effectively describe the hardest particles inside a jet. We get substructure information from N-subjettiness observables and (generalized) energy correlation functions.

To be more concrete, observables employed in this work for quark/gluon discrimination are as follows:
\begin{itemize}
\item Particle multiplicity and charged particle multiplicity of a jet.
\item Jet mass $m_J$, jet transverse momentum $p_{TJ}$ and their ratio $m_J/p_{T J}$.
\item Generalized angularities with two parameters $\kappa$ and $\beta$, which are proposed in \cite{Larkoski:2014pca} and further discussed in \cite{Gras:2017jty}:
\bea
\lambda_\beta^\kappa = \sum_{i\in J} z_i^\kappa \theta^\beta_i.
\eea
Using anti-$k_t$ algorithm \citep{Cacciari:2008gp} with E-scheme recombination \citep{Blazey:2000qt} in proton-proton collisions, one has
\bea\label{z_theta_GA}
z_i\equiv {p_{Ti}\over \sum_{j\in J} p_{Tj}},\quad\theta_i\equiv {R_{i\hat n}\over R_0},
\eea
where $z_i$ is the momentum fraction of particle $i$, $R_{i\hat n}$ is the angular distance between a chosen axis ${\hat n}$ and particle $i$, and $R_0$ is the radius of the jet under consideration.\footnote{We use  the axis directly from E-recombination in this paper. An alternative is with the winner-take-all recombination scheme \cite{Bertolini:2013iqa, Larkoski:2014uqa}.}
In particular, five sets of $(\kappa, \beta)$ from \cite{Larkoski:2014pca, Gras:2017jty} are used as benchmarks,
\bea
(0,\, 0),\qquad(2,\,0), \quad &(1, \, 0.5),&\quad (1, \, 1),\quad (1,\,2)\nonumber\\
\textrm{multiplicity}~~\quad p_T^D ~~\quad &\textrm{LHA}&~\quad \textrm{width}~~\quad \textrm{mass}\nonumber
\eea
Therein, observables with $\kappa=1$ are all IRC safe while the other two are IRC unsafe.
Each of them corresponds to a specific physical quantity: (1) $(0, \, 0)$ duplicates the jet multiplicity; (2) $(2,\, 0)$ is known as $p_T^D$ \cite{Chatrchyan:2012sn,Larkoski:2013eya}; (3) $(1, \, 0.5)$ is denoted as ``LHA" (Les Houches Angularity) \cite{Gras:2017jty}; (4) the width $(1,\,1)$ is related to broadening or girth \cite{Catani:1992jc, Rakow:1981qn, Ellis:1986ig}; (5) the mass $(1,\, 2)$ is related to thrust \cite{Farhi:1977sg}.

\item N-subjettiness observables \cite{Thaler:2010tr, Thaler:2011gf} measure the radiation about N selected axes in a jet with a definition of
\bea
\tau_N^{(\beta)}={1\over p_{T J}} \sum_{i\in J} p_{T\,i}\, \textrm{min}\left\{R^\beta_{1\,i}, R^\beta_{2\,i},\dots, R^\beta_{N\,i}\right\}.
\eea

It's demonstrated in \cite{Datta:2017rhs} that a basis can be constructed with $(3M-4)$ N-subjettiness observables to span the phase space of appropriately identified $M$ particles. This basis is then taken as the input of deep neural networks to discriminate the boosted hadronic Z decay from light parton initiated jets.
The N-subjettiness observables chosen in this work are list below,
\beq
\label{Nsub}
\left\{
\begin{array}{l}
\tau_1^{(0.5)},\tau_1^{(1)},\tau_1^{(2)},\tau_2^{(0.5)},\tau_2^{(1)},\tau_2^{(2)},\\ \tau_3^{(0.5)},\tau_3^{(1)},\tau_3^{(2)},\tau_4^{(0.5)},\tau_4^{(1)},\tau_4^{(2)},\tau_5^{(1)},\tau_5^{(2)}
\end{array}
\right\}.
\eeq

They are identical to the ones used in \cite{Datta:2017rhs, Datta:2017lxt} for spanning the 6-body phase space in a jet. Two ratios $\tau_{21}^{(1)}= \tau_2^{(1)}/ \tau_1^{(1)}$ and $\tau_{21}^{(2)}= \tau_2^{(2)}/ \tau_1^{(2)}$ are also included.
The ``OnePass$\_$WTA$\_$KT$\_$Axes" in the FastJet package is chosen in the calculation of N-subjettiness observables.

\item Generalized energy correlation functions \cite{Larkoski:2013eya, Moult:2016cvt} allow one to identify N-prong jet substructure without being required to find subjets at first place like for N-subjettiness. In this work, we employ $C_N^{(\beta)}$ with $N=1$ for quark/gluon discrimination \cite{Larkoski:2013eya} and $U_i$'s  \cite{Moult:2016cvt}. We briefly introduce these observables before moving forward.
The definition of $C_N^{(\beta)}$ \cite{Larkoski:2013eya} is as follows,
\bea
C_N^{(\beta)}\equiv{r_N^{(\beta)}\over r_{N-1}^{(\beta)}},\qquad
r_N^{(\beta)}\equiv{\textrm{ECF}(N+1, \beta)\over \textrm{ECF}(N, \beta) },
\eea
\bea
\textrm{ECF}(N, \beta)=\sum_{i_{a,b,c}\in J}\left(\prod_{a=1}^Np_{T i_a}\right)\left(\prod_{b=1}^{N-1}\prod_{c=b+1}^{N}R_{i_b i_c}\right)^\beta,
\eea
 where $r_N$ is much like the N-subjettiness $\tau_N$ while $C_N^{(\beta)}$ is similar to the N-subjettiness ratio $\tau_{N,N-1}^{(\beta)}=\tau_{N}^{(\beta)}/\tau_{N-1}^{(\beta)}$.
 Both $C_N^{(\beta)}$ and $\tau_{N,N-1}^{(\beta)}$ are effective to probe the N-prong substructure in a jet. In particular, they are good measures of higher-order radiation corrections to the leading-order description.
Observable $U_i$ is proposed in \cite{Moult:2016cvt} as
 \bea
U_i^{(\beta)}=~_1 e_{i+1}^{(\beta)},
\eea
where
\bea
_v e_n^{(\beta)}&=&\sum_{1\leq i_1<i_2<\dots<i_n\leq n_J} z_{i_1} z_{i_2}\dots z_{i_n}\times\prod_{m=1}^v \textrm{min}_{s<t\in \{i_1,i_2,\dots, i_n\}}^{(m)} \left\{\theta_{st}^\beta\right\}.
\eea
For $pp$ collisions, $z_i$'s share the same definition in \eqref{z_theta_GA}  and $\theta_{ij}$ denoting the opening angles between constituent $i$ and $j$ of a jet. It's shown in~\cite{Moult:2016cvt} that the $U_i^{(\beta)}$ series, which is able to count higher-point correlators, are powerful in the discrimination of light quark jets and gluon jets. For this reason, we include $C_1^{(0.5)}$, $C_1^{(0.2)}$ and six different $U_i^{(\beta)}$ objects with $i = 1, 2, 3$ and $\beta=0.5,\,0.2$ into our training data.
As a complement, three general energy correlators $\textrm{ECF}(2, \beta)$ with $\beta= 0.5,\,1.0$ and $2.0$ are also included.
\end{itemize}

\subsection{Database generation}

We simulate jet events produced from $p p$ collisions at the LHC with $\sqrt s =13$~TeV.
The database consists of observables from four different ranges of jet transverse momentum ($p_{T J}$): $[200, \, 220]$ GeV, $[500, \, 550]$ GeV, $[1000, \, 1100]$ GeV and $[200,\, 1000]$ GeV. The first three bins represent jets with $p_{T J}=200, 500$ and $1000~{\rm GeV}$.
The reason to include the last wide $p_{T J}$ range is to train a single classifier for jets in various kinematic regions, which is advantageous compared to the case of DCNN as we will discuss later.

We generate quark events from $p p \rightarrow  q q, q \bar{q}$ hard processes with MadGraph5 v2.5.5 \citep{Alwall:2014hca}, where $q$ includes only light quarks $(u,\,d,\, s)$.
Similarly, gluon events are generated from $p p \rightarrow g g $ process.
\footnote{It's worth specifying the definitions of quark and gluon jets.
As is discussed in \cite{Gras:2017jty}, these lead to ambiguities for jet tagging.
Since the processes considered in this study are enough simple and the radius of jet clustering is moderate, we take the straightforward definition, namely a one-to-one mapping between a jet and its initiating parton. Therefore, jets from $p p \rightarrow g g $ are denoted as gluon jets while those from $p p \rightarrow  q q, q \bar{q}$ are quark jets.}
Production channels for mixed quark and gluon final states like $p p \rightarrow q g$ have been shut down to avoid any ambiguity.
The transverse momentum cuts at parton level are $20\%$ broader than the $p_{T J}$ windows. All other cuts are turned off.
In this way, we are able to prevent most bias from kinematic cuts and in the meantime ensure the data generation efficiency.

We pass parton level events into PYTHIA 8.226 \cite{Sjostrand:2014zea} and shower them with default parameter settings.\footnote{It is known that gluon jets generated with PYTHIA could be quite different from those with older version HERWIG. However, a recent study on the thrust event shape in \cite{Mo:2017gzp} found the difference between PYTHIA and HERWIG 7.1 has become smaller for gluon jets at hadron level. Both results are consistent with precise analytic predictions and within their uncertainties. Therefore we only use PYTHIA for parton shower in this study.} In the showered events, only final state particles with $|\eta| < 2.5$ are kept while invisible neutrinos are discarded.
FastJet 3.3.0 \citep{Cacciari:2011ma} is used to cluster particles with anti-$k_T$ algorithm \citep{Cacciari:2008gp} and E-scheme recombination \citep{Blazey:2000qt}. We use $R_0=0.4$ for jet radius.
Jet mass and transverse momentum can be directly read out from FastJet.
Observables including charged particle multiplicity and five benchmark generalized angularities are calculated with our private codes based on FastJet. N-subjettiness observables and relevant energy correlation functions are all derived with FastJet-contrib 1.027.

In the end, we have one million available events in each narrow $p_{T J}$ region: $[200, \, 220]$ GeV, $[500, \, 550]$ GeV, $[1000, \, 1100]$ GeV and $1.3$ million events for the $[200, \, 1000]$ GeV case. Each sample consists of half quark jet events and half gluon jet events.
One should notice that the generation of $[200,\, 1000]$ GeV data requires more careful treatment for two reasons.
First, bias from kinematic cut becomes negligible in this situation.
Secondly, the parton level $p_{TJ}$ differential distribution drops quickly with $p_{TJ}$.
This indicates that it would be more efficient to generate events, not once-through, but in many sub-windows, with approximately equal numbers of events in each sub-window.
Therefore, we generate events in sixteen equal-width bins from $200$ to $600$ GeV and eight equal-width bins from $600$ to $1000$ GeV. The events gained from all twenty-four bins make up the broad kinematics region data for our following analysis.

In realistic discrimination work, detector effects would unavoidably reduce the performance of a classifier when soft and collinear jet features are not fully captured due to finite detector resolutions. A better understanding of such effects is also necessary for the use of machine learning techniques in future data-driven studies. The ATLAS collaboration already presented a study of large-radius jet observables including N-subjettiness and energy correlation functions from real data samples in \cite{TheATLAScollaboration:2015ynv}.
In the meantime, the authors of \cite{Tripathee:2017ybi, Larkoski:2017bvj} studied hard 2-prong substructure and QCD splitting functions from CMS Open Data and found a good agreement with particle-level simulations.
Concerning this systematic issue, we generated detector level data as a comparison. The detector simulation is performed with Delphes \cite{deFavereau:2013fsa} and the standard CMS detector card based on the Pythia output generated in the same way as in the particle level case. Then the energy-flow data is clustered by FastJet with the same settings for all observables.

\section{Architecture of fully connected neural networks}\label{sec:ArchitectureFNN}

\begin{figure*}[!htb]
  \begin{center}
    \includegraphics[width=0.9\textwidth]{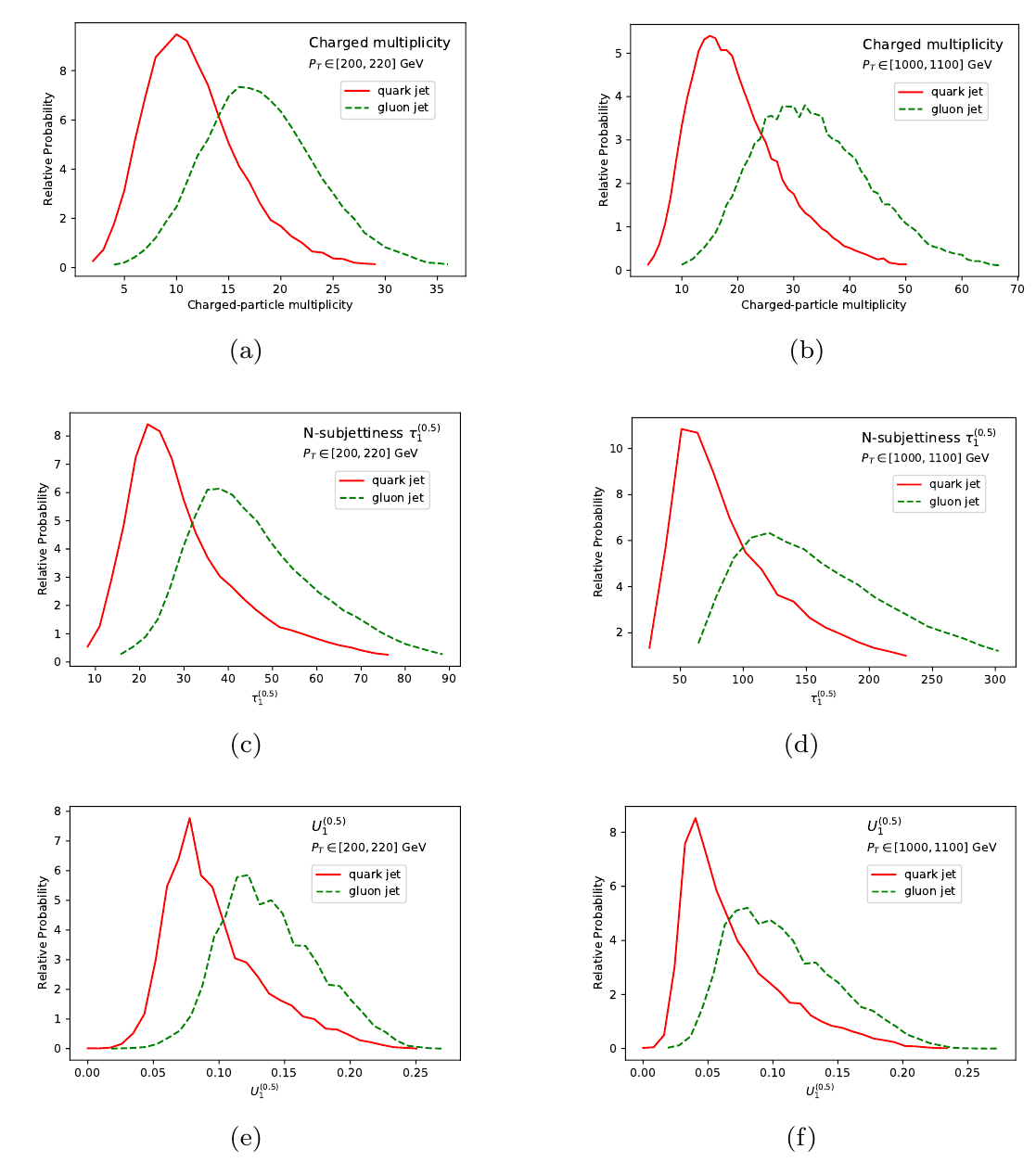}
  \end{center}
  \captionsetup{singlelinecheck = false, justification=raggedright}
  \caption{Distributions of three observables sensitive to quark/gluon jet tagging, measured on samples showered by Pythia in $p_{T J}$ bins of $[200, ~220]$ GeV (left) and $[1000, ~1100]$ GeV (right).
  From top to bottom are the distributions of charged particle multiplicity, N-subjettiness $\tau_1^{(0.5)}$ and energy correlation function $U_1^{(0.5)}$ for quark (red solid) and gluon jets (green dashed). The distributions are normalized to keep the $y$-axis range roughly in $0$-$10$.}
  \label{dist}
\end{figure*}

We have described in the last section how to generate simulation data containing $36$ expert-designed jet observables.
In Fig. \ref{dist}~(a)-(f), the distributions of charged particle multiplicity, N-subjettiness $\tau_1^{(0.5)}$ and energy correlation function $U_1^{(0.5)}$ are plotted as examples. These observables are sensitive to the quark/gluon tagging. The curves are obtained based on the two million events showered with Pythia in the $p_{TJ}$ bins of $[200, ~220]$ GeV and $[1000, ~1100]$ GeV respectively.

Before feeding these observables into the neural network, it is better to first
standardize them, so that the mean value and standard deviation of every feature are set to $0$ and $1$.
The one million events in each given $p_{TJ}$ range are split into three sets: $8\times 10^5$ events for training, $10^5$ events for validation and the rest $10^5$ events as test data.
We use a fully connected deep neural network with six hidden layers.
Each hidden layer has $300$ nodes.
For the activation function, we choose the rectified linear unit function (ReLU) for hidden layers and the sigmoid function for the output layer. The binary cross-entropy loss function is minimized using the RMSprop optimization \cite{RMSprop} with the initial learning rate $0.00015$ and the decay factor $0.995$ for the history gradient.

\begin{figure*}[!htb]
  \begin{center}
    \includegraphics[width=0.4\textwidth]{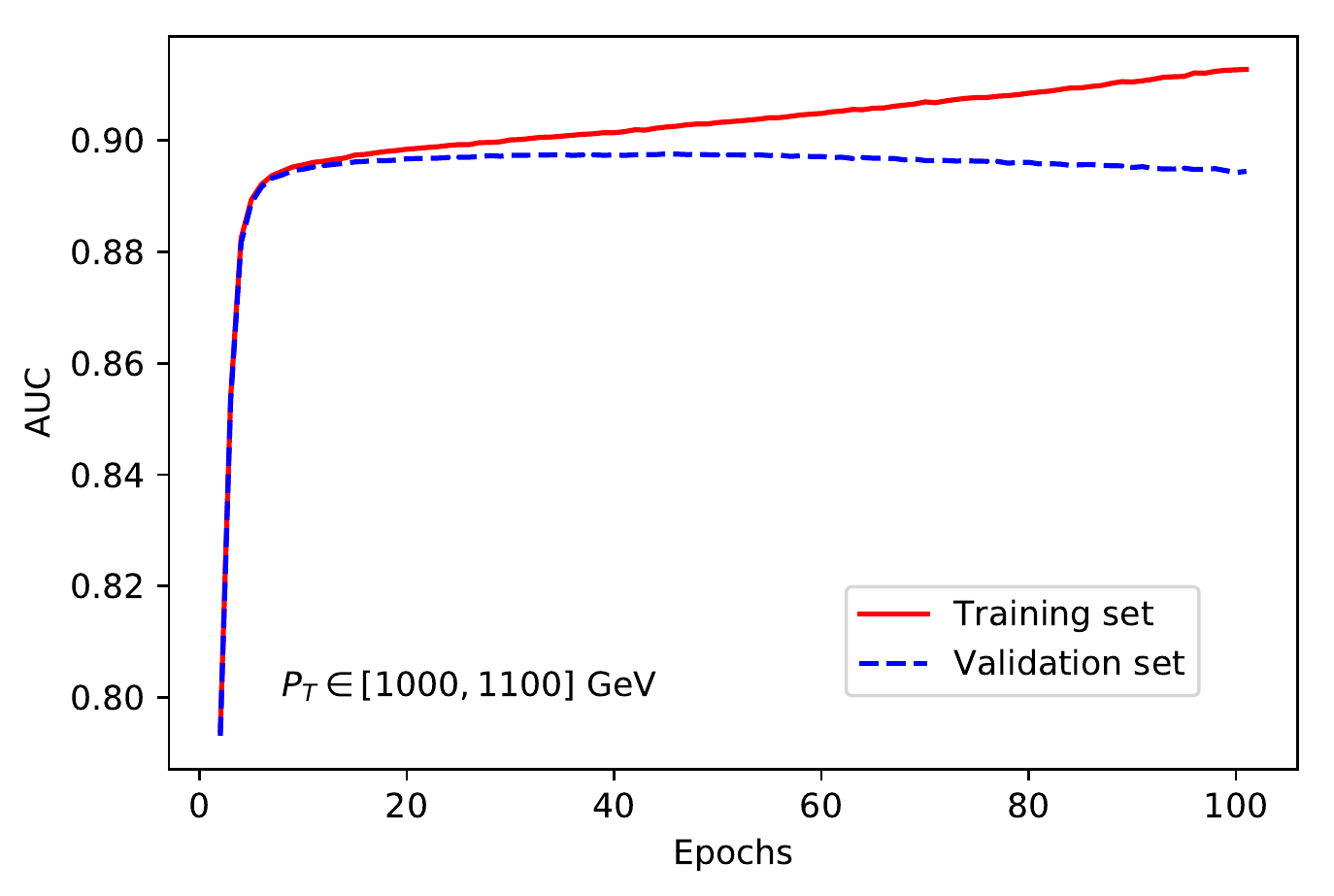}
  \end{center}
  \captionsetup{singlelinecheck = false, justification=raggedright}
  \caption{The AUC learning curves of the fully connected neural network trained on jets with their transverse momentum in the range of $[1000, 1100]$ GeV. }
  \label{learning-curve}
\end{figure*}

The model contains more than $460$k unknown weights and biases as parameters and it is important to prevent its overfitting.
For this purpose, the dropout regularization and validation-based early stop are adopted. The dropout ratio is taken to be $0.1$ for all six hidden layers. The learning curves of AUC vs. epochs are shown in Fig. \ref{learning-curve} as an illustration for light quark and gluon jets with the transverse momentum $p_{T J} \in [1000, 1100]$ GeV. One can see from the figure that the validation AUC starts to decrease slowly at around $40$ epochs, while the training AUC continues to increase with more training epochs. So we would stop the training if its performance on the validation data does not improve anymore for $20$ epochs. The neural network is initialized with random Gaussian weights of zero mean. The standard deviations are chosen to be inversely proportional to the square root of the number of nodes in the corresponding layer.

The neural network model is implemented with Tensorflow \cite{Tensorflow} and scikit-learn packages \cite{scikit}.
Running on an NVidia GTX 1080 GPU, it takes just several minutes to train the model with input events in one given $p_{TJ}$ region.

\section{Results}\label{sec:results}

\begin{figure*}[!htb]
  \begin{center}
    \includegraphics[width=0.9\textwidth]{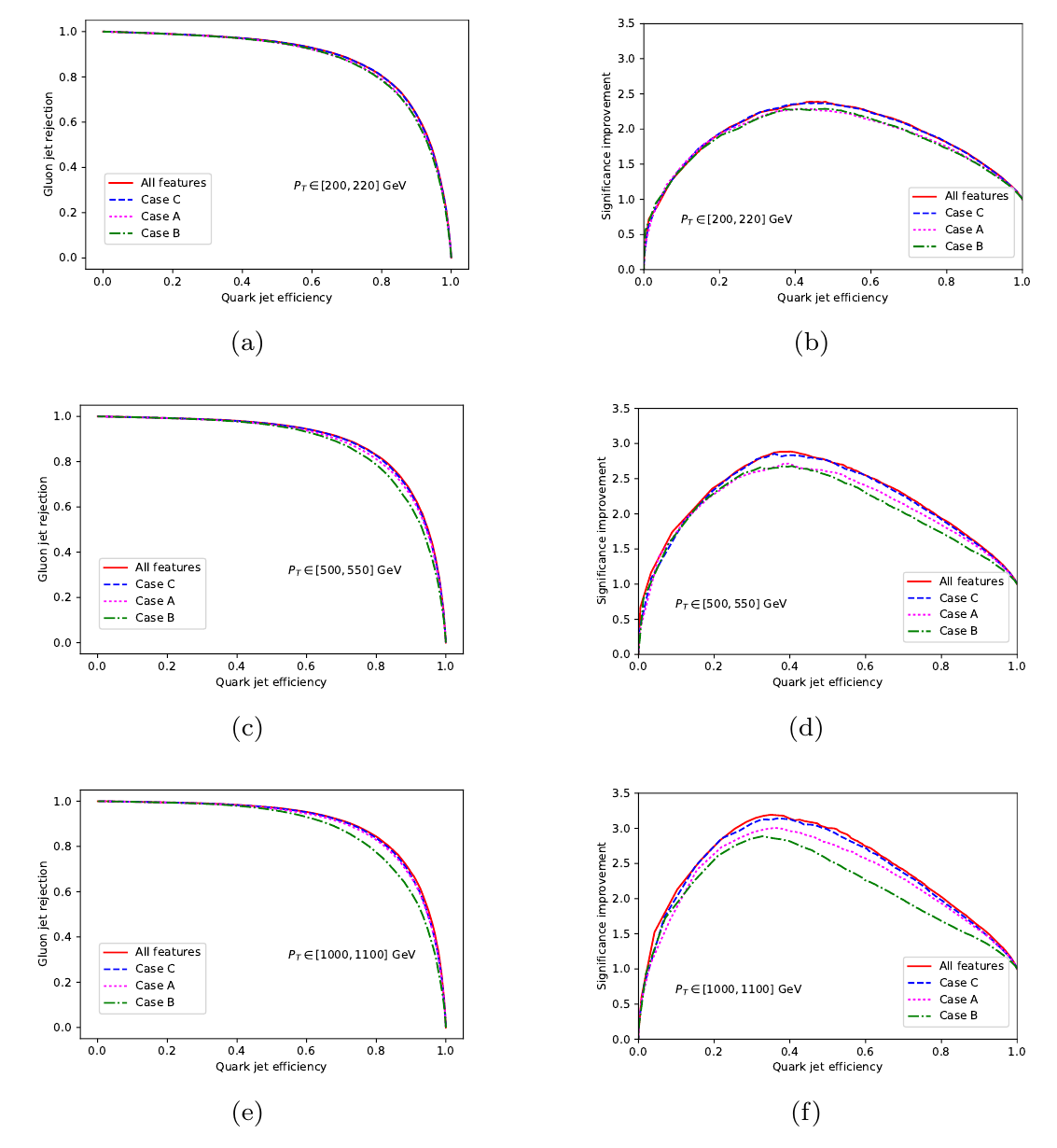}
  \end{center}
  \captionsetup{singlelinecheck = false, justification=raggedright}
  \caption{The ROC (left) and SIC (right) curves of fully connected neural networks trained on jets at particle level with their transverse momentum in given bins of $[200, 220]$ GeV (top), $[500, 550]$ GeV (middle) and $[1000, 1100]$ GeV (bottom). The red solid curves represent results using all jet observables as input features, while blue dashed, magenta dotted and green dot-dashed curves represent results using different subsets of jet observables, as explained in the context.}
  \label{ROC-sic}
\end{figure*}

In analog to new physics searches, we treat light quark jets as signals because they are presumably associated with new particles. The gluon jets are thus taken to be backgrounds.
Therefore, the performance of a neural network can be measured by gluon rejection efficiency ($1-\epsilon_g$) as a function of quark acceptance efficiency ($\epsilon_q$), known as the receiver operating characteristic (ROC) curve which is widely used in the field of machine learning.
The area under the ROC curve (AUC) is also a useful quantity to measure the performance of models.
Moreover, in collider physics, it might be better to plot the significance improvement characteristic (SIC) curve $\epsilon_q/\sqrt{\epsilon_g}$, which is directly related to the statistical significance of the signal and background separation.

In Fig. \ref{ROC-sic}~(a)-(f), red solid lines represent ROC and SIC curves of neural networks with all jet observables discussed in Section \ref{sec:observables-and-data-generation} as input.
The ROC AUCs are $0.877$, $0.891$ and $0.899$ for jets with transverse momentum $p_{TJ}$ in the range of $[200, 220]$ GeV, $[500, 550]$ GeV and $[1000, 1100]$ GeV, respectively.
Currently, the DCNN with color \cite{Komiske:2016rsd} has the best performance in quark/gluon discrimination. To compare with results from DCNNs with color, we show in Table \ref{gluon-eff} the gluon jet efficiency at $50\%$ quark jet acceptance.\footnote{Our results can be directly compared with those of Ref. \cite{Komiske:2016rsd} because our events were generated purposely in much the same way as in Ref. \cite{Komiske:2016rsd}. Briefly, both showered event samples were simulated in $pp$ collisions at $\sqrt{s}=13$ TeV using PYTHIA. Then FastJet was used to cluster particles in the pseudorapidity range of $|\eta | < 2.5$ into jets using the anti-$k_T$ algorithm with the jet radius set to $0.4$. As far as we know, the only difference is that we used MadGraph5 first to generate parton level events which were then passed into PYTHIA for hadronic shower, while in \cite{Komiske:2016rsd} only PYTHIA was used for event generation.}
It turns out that our results from fully connected neural networks are basically as well as those of DCNNs with color for $p_{T J}$ around $200$ GeV. For larger $p_{TJ}$, for example around $1000$ GeV, our results are even slightly better than those from DCNNs with color.\footnote{This may due to the pixel-limited input of jet image CNN. The pixel sizes acts as a kind of coarse-graining, which might affect the efficiency of the CNN especially in the high transverse momentum region. The image size dependence was discussed in Ref. \cite{Komiske:2016rsd} only for $200$ GeV jets. Compared to the standard $33 \times 33$ grid size, a decrease of performance is observed for $13 \times 13$ pixelization. However the result of $43 \times 43$ pixelization is also slightly worse than that of $33 \times 33$ pixelization. It should be interesting to investigate further on the image size dependence for higher transverse momentum jets and/or larger grid sizes.}
This indicates that almost all information of quark/gluon jets has been included in jet observables used in this study.

As an attempt to figure out which jet observables are more important for the quark/gluon discrimination, we also train neural networks by using the following different subsets of jet observables:
 \begin{itemize}
   \item Case A. Input features include jet mass $m_J$, fourteen N-subjettiness observables $\tau_N^{(\beta)}$ listed in Eq. \ref{Nsub} and three ratios $m_J/p_{TJ}$, $\tau_2^{(1)}/\tau_1^{(1)}$ and $\tau_2^{(2)}/\tau_1^{(2)}$.
   This basis is essentially the same as the one proposed in \cite{Datta:2017rhs}, which is powerful in the discrimination of boosted Z jets from QCD backgrounds.
   The results are shown in Fig. \ref{ROC-sic} as magenta dotted curves.
   One can directly compare them with the red solid curves with all jet observables. The difference appears very small using ROC measures, while the difference becomes more and more obvious in SIC curves for jets with larger transverse momentum.
   \item Case B. This subset first includes jet mass, particle multiplicity, generalized angularities $\lambda_\beta^\kappa$ with $(\kappa, \beta)$ equals to $(2,0)$,  $(1,0.5)$, $(1,1)$ and $(1,2)$.
   In addition, energy correlation functions ECF$(2,\beta)$ with $\beta=0.5, 1, 2$ and ratios $C_1^{(0.2)}$ and $C_1^{(0.5)}$ are involved.
   It was found in \cite{Moult:2016cvt} that a new set of energy correlation functions $U_i^{(\beta)}$ are powerful for the quark/gluon tagging.
   So six $U_i^{(\beta)}$ observables ($i=1,2,3$ and $\beta=0.2, 0.5$) are also taken into account here.
   The results are shown in Fig. \ref{ROC-sic} as green dot-dashed curves.
   However, this case has somehow inferior performance compared to other cases.
   \item Case C. Fourteen jet observables are considered, which includes particle multiplicity, charged particle multiplicity, LHA $\lambda_1^{(0.5)}$, jet mass, energy correlation function $U_1^{(0.5)}$ and nine N-subjettiness observables $(\tau_1^{(0.5)},\tau_1^{(1)},\tau_1^{(2)},\tau_2^{(0.5)},\tau_2^{(1)},\tau_2^{(2)},\tau_3^{(0.5)},\tau_3^{(1)}, \tau_3^{(2)} )$.
       The results are shown in Fig. \ref{ROC-sic} as blue dashed curves.
       Interestingly, these results are basically as good as those using all jet observables.
       This implies that these fourteen observables may have captured most physical information that is useful for discriminating quark/gluon jets.
 \end{itemize}

\begin{figure*}[!htb]
  \begin{center}
    \includegraphics[width=1\textwidth]{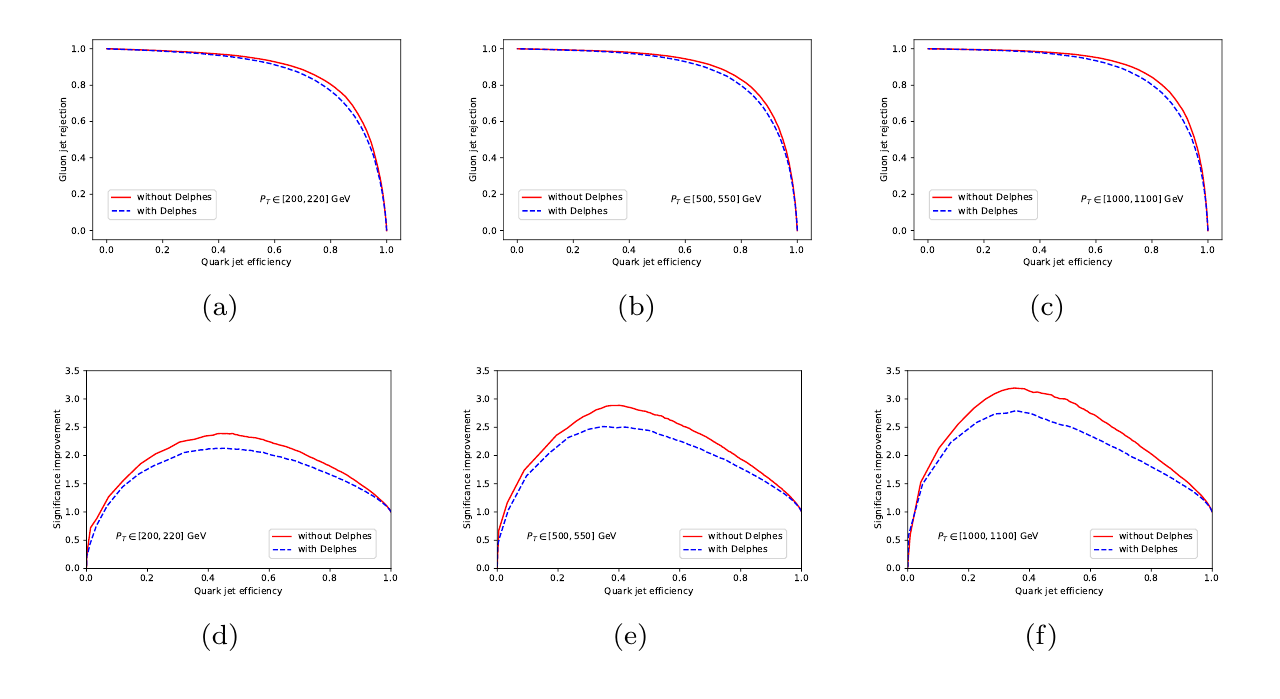}
  \end{center}
  \captionsetup{singlelinecheck = false, justification=raggedright}
  \caption{Comparison of the ROC (top) and SIC (bottom) curves with (blue dashed) or without (red solid) detector effects. The corresponding transverse momentum are in the range of $[200, 220]$ GeV (left), $[500, 550]$ GeV (middle) and $[1000, 1100]$ GeV (right), respectively. }
  \label{Delphes}
\end{figure*}

When detector effects are taken into account, the classification quality of quark/gluon jets would be worsen due to finite resolutions in real measurements. Concerning this issue, we also perform a fast detector simulation with Delphes employing standard CMS detector card based on the Pythia output. Then particles are clustered using FastJet with same settings in the particle level case. The corresponding ROC curves are plotted in Fig. \ref{Delphes}~(a)-(c) and SIC curves in Fig. \ref{Delphes}~(d)-(f). The ROC AUCs at the detector level are $0.861$, $0.877$ and $0.881$, respectively, for jets in three transverse momentum bins. The difference of ROC AUCs with or without detector effects is about $0.015$. For SIC curves, one can see from the figure that impacts of finite resolutions are less than about $15 \%$ for jets with different transverse momenta.

Notice that jet images look different for various transverse momenta, it is usual to train as many DCNNs as the number of selected transverse momentum benchmark regions.
For example, three DCNNs are trained in \cite{Komiske:2016rsd} to classify quark/gluon jets with $p_{TJ} \in [200, 220]$ GeV, $[500, 550]$ GeV and $[1000, 1100]$ GeV.
However, the classification quality may suffer some loss for jets with $p_{TJ}$ falling outside the considered bins when using any of these three well-trained DCNNs. This is clearly not the most efficient method.
Up to now, we are strictly following the way of DCNN such that different FNNs are trained for various jet transverse momentum regions.
However, since transverse momentum is just one of the input features of FNN, we should in principle be able to train a single FNN for jets with very different transverse momentum.

\begin{figure*}[!htb]
  \begin{center}
    \includegraphics[width=0.9\textwidth]{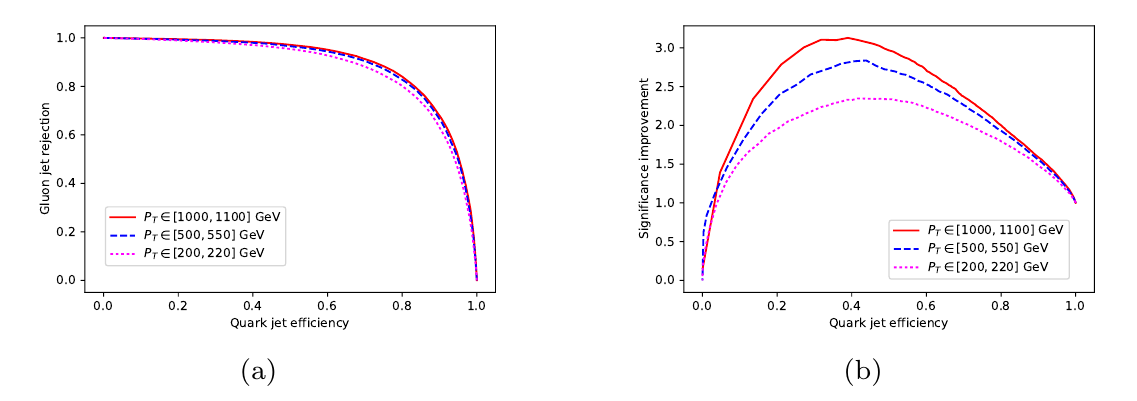}
  \end{center}
  \captionsetup{singlelinecheck = false, justification=raggedright}
  \caption{The ROC (left) and SIC (right) curves which are trained on jets with the transverse momentum in the range of $[200, 1000]$ GeV while tested on jets with transverse momentum in the bins of $[200, 220]$ GeV (magenta dotted lines), $[500, 550]$ GeV (blue dashed lines) and $[1000, 1100]$ GeV (red solid lines). }
  \label{ROC200-1000}
\end{figure*}

As an attempt, a single FNN is trained for quark/gluon jets with $p_{TJ}$ in the range of $[200, 1000]$ GeV.
We then test this FNN on jets with $p_{TJ}$ in $[200, 220]$ GeV, $[500, 550]$ GeV and $[1000, 1100]$ GeV.
The corresponding ROC and SIC curves are shown in Fig. \ref{ROC200-1000}~(a) and (b), with ROC AUCs being $0.876$, $0.890$ and $0.897$ for  $p_{TJ}$ in $[200, 220]$ GeV, $[500, 550]$ GeV and $[1000, 1100]$ GeV, respectively.
Comparing Figs. \ref{ROC-sic} and \ref{ROC200-1000}, one can see that a single FNN can discriminate quark/gluon jets in a wide range of transverse momentum without loosing any visible efficiency.

To compare our results from FNNs in a quantitative way with those from DCNNs with color \cite{Komiske:2016rsd}, the gluon jet efficiencies at $50\%$ quark jet acceptance are shown in Table \ref{gluon-eff}.
For $p_{TJ}$ around $200$ GeV, the performance of FNN is comparable to that of DCNN with color.
As $p_{TJ}$ increases, the performance of FNN becomes even better than that of DCNN with color.

\begin{table}
\begin{tabular}{|l|c|c|c|}
\hline
Gluon jet efficiency at & $1000$ GeV & $500$ GeV & $200$ GeV \\
$50\%$ quark jet acceptance & ($\%$) & ($\%$) & ($\%$) \\
\hline
\hline
FNN using all jet observables & $2.8$ & $3.3$ & $4.5$ \\
FNN with Case A & $3.2$  & $3.7$ & $5.0$ \\
FNN with Case B & $3.8$  & $3.9$ & $4.8$ \\
FNN with Case C & $2.8$  & $3.3$ & $4.6$ \\
\hline
\hline
A single FNN trained using  & & & \\
jets with $p_{TJ} \in [200, 1000]$ GeV & $2.8$  & $3.4$ & $4.6$ \\
\hline\hline
DCNN with color \cite{Komiske:2016rsd} & $3.4$ &--- & $4.6$ \\
\hline
\end{tabular}
\captionsetup{singlelinecheck = false, justification=raggedright}
\caption{Gluon jet efficiency at $50\%$ quark jet acceptance for the transverse momentum of jets in the range of $[200, 220]$ GeV, $[500, 550]$ GeV and $[1000, 1100]$ GeV, respectively.}
\label{gluon-eff}
\end{table}

As a comparison, we also run a shallow neural network containing only a single hidden layer with 300 nodes. The performance of the shallow neural network drops very slowly with decreasing numbers of nodes. For example, when the number of hidden nodes reduces from 300 to 50, the decrease in AUC is only about 0.001 for jets with $p_{T J} \in [1000, 1100]$ GeV. 
All the other hyper-parameters are chosen to be the same as those of deep neural networks. Since deeper neural network can express more complicated nonlinear functions, it has been shown in many situations that deep neural networks may provide a significant boost in the performance compared to shallow ones.  But surprisingly, in our case, we find that the performance of the shallow neural network is just as good as the deep one, though a larger number of epochs is required to train the shallow network. We do not plot the ROC and SIC curves of the shallow neural network, because they overlap completely with those of the deep neural networks. The fact that shallow network is already good enough may indicate that it is not very difficult to extract information from these jet observables. If this were true, shallow network may also have good performance in general jet classification. We have checked explicitly that this is indeed the case for the separation of the boosted Z jets from QCD jets.\footnote{We have generated the events following the procedure outlined in Ref.\cite{Datta:2017rhs}.} It should be interesting to check the power of shallow neural network in other situations, with these jet observables or other appropriate features as input.

\section{Summary}\label{sec:discussion}
Deep learning approaches have developed many applications in high energy physics, among which is jet identification, such as the separation of quark-initiated jets from gluon-initiated jets.
A conventional way is to take the energy deposited in the calorimeter as jet images.
As a powerful tool, deep convolutional neural networks can then be used to classify jet images.

Motivated by \cite{Datta:2017rhs}, we take $36$ expert-designed jet observables as input features in this paper to discriminate quark/gluon jets using fully connected neural networks.
One advantage of this method is that, the architecture of fully connected neural networks is much simpler than that of convolutional neural networks, and the former is also less GPU time-consuming.
Since jet images do not lose any information, the approach of convolutional neural networks may be more powerful in jet classification if (nearly) all of the useful information could be extracted from the data.
However, only a few pixels are activated in a jet image and it remains an open question whether the convolutional neural network is the most efficient method in this situation.

We first use a neural network with six hidden layers where $300$ nodes are set in each hidden layer.
Three neural networks are trained separately, based on one million events for each jet transverse momentum bins of $[200,~220]$ GeV, $[500,~550]$ GeV and $[1000,~1100]$ GeV.
As expected, the larger the jet transverse momentum, the better the performance of the neural networks.
Specifically, the ROC AUCs are $0.877$, $0.891$ and $0.899$ for the above three transverse momentum bins (larger AUC usually means better performance).
The gluon jet efficiencies at $50\%$ quark jet acceptance are $4.5\%$, $3.3\%$ and $2.8\%$ for the three transverse momentum bins.
These results are comparable to those from convolutional neural network for $p_{TJ} \in [200, ~220]$ GeV, and even slightly better than those from convolutional neural network for $p_{TJ} \in [1000, ~1100]$ GeV.

Many of these $36$ jet observables should be complementary for the quark/gluon tagging, but some of them may be redundant.
As an attempt, we test the performance of neural networks by choosing different subsets of jet observables as input features.
It is interesting to see that the neural network using only fourteen observables has as good performance as the neural network using all of thirty-six jet observables. These fourteen observables include particle multiplicity, charged particle multiplicity, LHA $\lambda_1^{(0.5)}$, jet mass, energy correlation function $U_1^{(0.5)}$ and nine N-subjettiness observables $(\tau_1^{(0.5)},\tau_1^{(1)},\tau_1^{(2)},\tau_2^{(0.5)},\tau_2^{(1)},\tau_2^{(2)},\tau_3^{(0.5)},\tau_3^{(1)}, \tau_3^{(2)} )$.

Since quark/gluon jet images may have different characteristics with various jet transverse momentum, it is customary to train different convolutional neural networks for different transverse momentum bins.
But $p_{TJ}$ is just one of the input features of FNNs, it should be possible to train a single FNN for all jet momentum regions. Therefore we train such a FNN using $1.3$ million data with jet transverse momenta in the range of $[200, ~1000]$ GeV. Again, the performance of such a FNN is almost the same as those of FNNs trained separately on each transverse momentum bins.

The above results do not take into account detector effects, which should worsen the classification accuracy due to finite resolution and limited acceptance. We estimate detector effects in quark/gluon separation by including a fast detector simulation with Delphes. The impacts on the statistical significance of the quark and gluon separation are roughly less than $15 \%$.

In general, deep neural network is expected to be more powerful than shallow network. We test the performance of a shallow neural network with one hidden layer of $300$ neurons. Surprisingly, this shallow neural network classifies light quark/gluon jets as well as the deep one, which implies that it is not very difficult to extract information from the jet observables we choose. Adopting  the same set of jet observables in \cite{Datta:2017rhs}, we find again that the shallow neural network gives approximately the same classification accuracy as the deep neural network for the separation of boosted Z jets and QCD backgrounds. It should be interesting to check whether performances of shallow neural networks could be as good as the deep ones in other situations.

In \cite{Datta:2017lxt}, a new method was proposed to construct novel jet observables with the help of fully connected neural networks. Such novel observables may have better discrimination power than widely-used observables. Novel observables may also deepen our understanding of the jet identification problem. It should be interesting to see in the future whether novel observables could also be constructed in the case of quark/gluon tagging.

\section*{Acknowledgement}
We gratefully thank Andrew Larkoski for helpful discussions, and thank Gavin Salam and Torbj\"{o}rn Sjostrand for clarifying technical details.
HL is supported by the German Science Foundation (DFG) within the Collaborative Research Center 676 ``Particles, Strings and the Early Universe'' and the Recruitment Program of Global Youth Experts of China.
Other authors are supported in part by the National Science Foundation of China (11135006,  11275168, 11422544, 11375151, 11535002) and the Zhejiang University Fundamental Research Funds for the Central Universities (2017QNA3007).

\end{document}